\documentclass[aps,twocolumn,prd,showpacs,showkeys,preprintnumbers,superscriptaddress,nobibnotes,floatfix,nofootinbib]{revtex4-2}

\usepackage{orcidlink}
\usepackage{amsmath}
\usepackage{amsfonts}
\usepackage{amssymb}
\usepackage{mathrsfs}
\usepackage{bm} 
\usepackage{bbm} 
\usepackage{slashed} 
\usepackage{physics} 
\usepackage{esvect} 
\usepackage[version=4]{mhchem}
\usepackage{xcolor} 
\usepackage{graphicx} %
\usepackage{array} 
\usepackage{url} %
\usepackage{cancel}
\usepackage{amsthm} 
	\usepackage{enumerate} 
		\usepackage{enumitem} %
		\usepackage{booktabs} 
		\usepackage{siunitx} 
		\usepackage{mdframed} %
		\usepackage{tikz}
		\usetikzlibrary{positioning}
		\usepackage[compat=1.1.0]{tikz-feynman}
		\usepackage{tcolorbox} %
		\usepackage{fontawesome5} 
		\usepackage{braket} 
		\usepackage{setspace} %
		\usepackage{multirow} %
		\usepackage{upgreek} 
		\usepackage{xspace} %
		\usepackage[normalem]{ulem} 
		\usepackage{array}
		\usepackage{natbib} 
		\usepackage{hyperref} %
		\hypersetup{
			colorlinks=true,
			linkcolor=blue,
			citecolor=blue,
			filecolor=magenta,
			urlcolor=blue,
			breaklinks=true
		}
		\usepackage[all]{hypcap} 
		\usepackage[capitalise]{cleveref} %
		\usepackage[normalem]{ulem}
		\usepackage{xcolor}

		\setlist[description]{leftmargin=\parindent,labelindent=\parindent} %

		\usepackage{slashed}

\begin{document}

			\title{Inelastic Scattering Effects on Attenuation of Boosted Dark Matter}
			
			\author{Guanhua Gu}
			\email{guguanhua@itp.ac.cn}
			\affiliation{Department of Physics and Institute of Theoretical Physics,
				Nanjing Normal University, Nanjing, 210023, P. R. China}
			\affiliation{Institute of Theoretical Physics, Chinese Academy of Sciences, Beijing 100190, P. R. China}
			\affiliation{School of Physical Sciences, University of Chinese Academy of Sciences, Beijing 100049, P. R. China}
			
			\author{Liangliang Su}
			\email{liangliang.su@kit.edu}
			\affiliation{Institute for Astroparticle Physics (IAP), Karlsruhe Institute of Technology (KIT),
				Hermann-von-Helmholtz-Platz 1, 76344 Eggenstein-Leopoldshafen, Germany}
			
			\author{Lei Wu}
			\email{leiwu@njnu.edu.cn}
			\affiliation{Department of Physics and Institute of Theoretical Physics,
				Nanjing Normal University, Nanjing, 210023, P. R. China}
			\affiliation{Nanjing Key Laboratory of Particle Physics and Astrophysics, Nanjing, 210023, China}
			
			\author{Jin Min Yang}
			\email{jmyang@itp.ac.cn}
			\affiliation{Centre for Theoretical Physics, Henan Normal University, Xinxiang 453007, P. R. China}
			\affiliation{Institute of Theoretical Physics, Chinese Academy of Sciences, Beijing 100190, P. R. China}
			
			\date{\today}
			
			\begin{abstract}
				Earth attenuation is crucial for interpreting direct-detection constraints on boosted dark matter (DM), since scatterings with terrestrial nuclei can significantly modify the flux and energy spectrum reaching underground detectors. At boosted energies, inelastic nuclear channels beyond ordinary elastic scattering can become relevant, including quasi-elastic scattering, deep-inelastic scattering, and resonant scattering. In this work, we incorporate the resonant scattering of boosted dark matter (DM) off nuclei into the Earth-attenuation framework, in combination with the elastic, quasi-elastic, and deep-inelastic channels. We find that, in the heavy-mediator regime, resonant scattering can give a non-negligible contribution to the attenuation of boosted DM. Using the latest PandaX-4T data, we derive new constraints on the spin-independent boosted DM-nucleon cross section $\bar{\sigma}_n$.
			\end{abstract}
			
			\maketitle
			
			\section{introduction}
			The existence of dark matter (DM) is strongly supported by a wide range of cosmological and astrophysical observations, and considerable effort has been devoted to understanding its nature. Among the various DM candidates, weakly interacting massive particles (WIMPs) have been particularly popular, owing to their natural connection to electroweak‑scale freeze‑out~\cite{Lee:1977ua,Roszkowski:2017nbc}. Although direct-detection experiments have achieved increasingly high sensitivity in searches for WIMPs~\cite{LZ:2024zvo,XENON:2025vwd,PandaX:2024qfu}, recent years have seen growing interest in dark-sector scenarios where the detectable component of dark matter is much lighter, typically in the keV--GeV mass range~\cite{Essig:2011nj,Hochberg:2015pha,Essig:2015cda,Essig:2017kqs,Knapen:2017xzo,DAgnolo:2018wcn,Wang:2019jtk,Ge:2020yuf,Guo:2020oum,PandaX-II:2021lap,Su:2021jvk,Calabrese:2021src,Bell:2021xff,Elor:2021swj,Alvey:2022pad,Su:2022wpj,PandaX:2023tfq,Liang:2024xcx,Dutta:2024kuj,Bhattiprolu:2024dmh,Sun:2025gyj,Balan:2025uke,Gong:2025ves,Cheek:2025nul,Ge:2025itf,Bernreuther:2025xqk,Hu:2025dsv,Cox:2025toz,Wang:2026you,Gong:2026dte}.
			
			However, the light mass that makes these candidates theoretically attractive also poses a serious experimental challenge for direct detection. For nonrelativistic light DM, the nuclear recoil energy is typically too small to exceed the detection threshold. This kinematic limitation becomes less severe for relativistic incoming DM, which has motivated growing interest in accelerated light-DM scenarios, including cosmic-ray up-scattered DM (CRDM)\cite{Yin:2018yjn,Bringmann:2018cvk,Xia:2020apm,Ema:2020ulo,Feng:2021hyz,Wang:2021nbf,PandaX-II:2021kai,Super-Kamiokande:2022ncz,Nagao:2022azp,Maity:2022exk,Lu:2023aar,Ghosh:2024dqw,LZ:2025iaw,Jeesun:2026lro}, atmospheric DM (ADM)\cite{Alvey:2019zaa,Su:2020zny,Arguelles:2022fqq,Darme:2022bew,Du:2022hms}, and boosted DM (BDM)\cite{Agashe:2014yua,Berger:2014sqa,Agashe:2015xkj,Super-Kamiokande:2017dch,Wang:2021jic,Granelli:2022ysi,Su:2023zgr,DeMarchi:2025uoo,Jeesun:2025gzt}. Constraints derived from the bare flux of these energetic DM particles only translate into a lower bound on $\bar{\sigma}_n$. In realistic scenarios, however, these particles must propagate through the Earth before reaching terrestrial detectors\cite{Xing:2026bbq}. For sufficiently strong attenuation, the particles can no longer deposit a detectable amount of energy in the detector. As a result, attenuation effects give rise to an upper boundary on $\bar{\sigma}_n$  in the final exclusion region.  Moreover, relative to the unattenuated flux, the DM flux arriving at the detector can be substantially modified by repeated scatterings with terrestrial nuclei. A detailed treatment of attenuation is therefore crucial for deriving reliable constraints.
			
			\begin{figure*}
				\centering
				\includegraphics[width=0.95\textwidth]{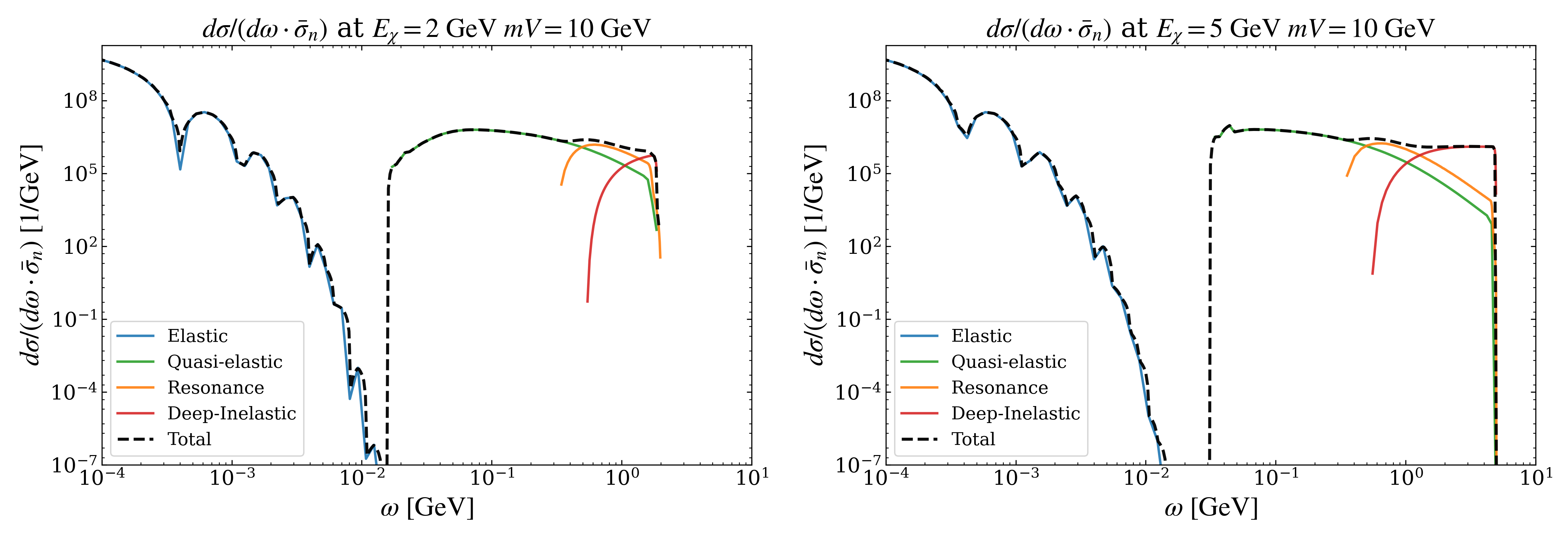}
				\vspace{-.2cm}
				\caption{
					Comparison of the normalized differential scattering cross sections for Fe,  with $m_V = 10~\mathrm{GeV}$ and $m_\chi = 1~\mathrm{MeV}$. Here $\bar\sigma_n$ is the reference DM--nucleon cross section used for normalization. The left and right panels correspond to incoming DM energies $E_\chi = 2~\mathrm{GeV}$ and $5~\mathrm{GeV}$, respectively. The blue, green, orange, red, and black dashed curves denote the elastic(ES), quasi-elastic (QES), resonance (RES), deep inelastic scattering (DIS), and total contributions, respectively.}
				\label{fig:dXS_comparison}
			\end{figure*}
			
			When interactions between DM particles and nuclei are considered, the finite size and internal structure of nuclei lead to different dominant scattering channels in different momentum-transfer regimes\cite{Soper:2014ska}, as can be clearly observed in Fig.\ref{fig:dXS_comparison}. Since the spatial resolution of the probe is set by the momentum transfer, $\lambda \sim 1/q$, the nucleus behaves coherently as a whole for $qR_A \lesssim 1$, where $R_A$ is the nuclear radius. In this regime, elastic scattering provides the dominant contribution to the total interaction rate. For larger momentum transfers, the probe begins to resolve individual nucleons, and the scattering becomes incoherent. The DM particle can then scatter from and eject a single nucleon from the nucleus, corresponding to quasi-elastic scattering. At still higher momentum transfers, the probe can resolve the partonic structure of nucleons and scatter off individual quarks. Since such large momentum transfers are difficult to achieve for ordinary low-energy DM, deep-inelastic scattering is usually negligible in conventional direct-detection settings. However, in extreme astrophysical environments, DM particles can be accelerated to much higher energies, making the DIS contribution important in corresponding  calculations~\cite{Su:2024flx,Li:2025zwg}.
			
			Between the quasi-elastic and deep-inelastic regimes, there exists a resonance region in which the nucleon is excited into a baryonic resonant state. Compared with quasi-elastic scattering, resonance production requires additional energy to excite the nucleon from its ground state to a heavier baryonic state, characterized by a mass gap $\Delta M = M_\Delta - m_N$. Therefore, the incident DM energy at which the resonance contribution becomes important is shifted to higher values relative to the quasi-elastic case, which coincides well with the typical energy range of accelerated DM, making resonance production an especially important channel in relevant calculations.
			
			In this work, we perform a systematic calculation of the resonance contribution to dark matter (DM)–nucleus scattering and adopt boosted dark matter (BDM) as a representative accelerated-DM scenario to examine how resonant interactions affect its attenuation. The remainder of this paper is organized as follows. Sec.~II describes the production mechanism of BDM and presents its bare flux. Sec.~III introduces the resonant scattering process and formulates its calculation within the impulse approximation\cite{Benhar:2005dj,Ankowski:2013gha,Su:2022wpj,Su:2023zgr}. Sec.~IV addresses the attenuation of BDM, incorporating two models of terrestrial attenuation into the transport analysis. Building on these results, Sec.~V evaluates the BDM flux at the PandaX-4T Run1 site, and derives the corresponding exclusion limits. Finally, Sec.~VI provides a summary.
			
			\section{Flux of Boosted Dark Matter}
			The BDM scenario describes a two-component ($\chi_1$ and $\chi_2$) DM model, with $\chi_1$ heavier than $\chi_2$. This model assumes that the dominant non-relativistic DM component, the heavy species $\chi_1$, does not directly interact with ordinary matter. However, $\chi_1$ can annihilate into the lighter state $\chi_2$ to produce the correct thermal relic abundance,
			\begin{equation}
				\chi_1 + \bar{\chi}_1\to\chi_2+\bar{\chi}_2,
			\end{equation}
			where $\chi_2$ is boosted to a relativistic state, with a Lorentz factor $\gamma_b = m_{\chi_1}/m_{\chi_2}$ determined by the mass ratio of the two DM components.  Here, $m_{\chi_1}$ and $m_{\chi_2}$ denote the masses of $\chi_1$ and $\chi_2$, respectively. In this work, we refer to the relativistic $\chi_2$ component as BDM, which can be detected through its interactions with Standard Model particles. 
			
			Because the annihilation rate scales as the square of the $\chi_1$ density, the dominant contribution to the $\chi_2$ flux at the Earth originates from the Galactic Center, where the DM density is highest. The differential flux from the galaxy can be written as \cite{Su:2023zgr}
			\begin{equation} \frac{\mathrm{d}\Phi_{\mathrm{\chi}}}{\mathrm{d}E_{\chi_2}}=\frac{1}{16\pi}\frac{\left<\sigma\nu\right>}{m^2_{\chi_1}}J(\Omega)\frac{\mathrm{d}N_{\chi_2}}{\mathrm{d}E_{\chi_2}}
				\label{eq:source_flux}
			\end{equation}
			where $\langle\sigma v\rangle$ is the thermally averaged annihilation cross section. In the $s$-wave limit, it is given by~\cite{Agashe:2014yua}
			\begin{equation}
				\langle \sigma v \rangle
				\simeq
				\frac{m^2_{\chi_2}}{8\pi \Lambda^4}(1+\gamma_b)^2
				\sqrt{1 - \frac{1}{\gamma_b^2}} ,
			\end{equation}
			where the parameter $\Lambda = 250~{\rm GeV}$ is set in this work. In this framework, a pair of $\chi_1$ annihilate into two mono-energetic $\chi_2$ with energy $m_{\chi_1}$. Thus, the differential energy spectrum $\mathrm{d}N_{\chi_2}/\mathrm{d}E_{\chi_2}$ per annihilation is a Direct function, but we adopt the approximate Gaussian form in this work,
			\begin{equation}
				\frac{dN_{\chi_2}}{dE_{\chi_2}}\approx\frac{2}{\sqrt{2\pi}\sigma_0 m_{\chi_1}}\exp\left(-\frac{(E_{\chi_2}-m_{\chi_1})^2}{2\sigma_0^2m^2_{\chi_1}}\right)
				\label{eq:DES}
			\end{equation}
			Here, $\sigma_0$ parametrizes the width of the fractional deviation of the BDM energy $E_{\chi_2}$ from its monochromatic value $m_{\chi_1}$.
				The precise value of $\sigma_0$ does not appreciably affect our final
				constraints as long as the injection spectrum remains sufficiently narrow
				compared with the energy range relevant for detection. In this regime, the
				constraints are mainly controlled by the attenuation-induced shift and
				suppression of the BDM flux rather than by the precise initial spectral width.
				We therefore adopt $\sigma_0=0.05$\cite{Salazar-Gallegos:2025aqn} as a
				benchmark value in this work.
			
			The $J$-factor in Eq.~(\ref{eq:source_flux}) is given by
			\begin{equation}
				J(\Delta\Omega)
				=
				\int_{\Delta\Omega} d\Omega
				\int_{0}^{\infty}
				\rho_\chi^2\!\big(r(r_\odot,\theta,s)\big)\, ds ,
			\end{equation}
			where $r(r_\odot, s, \theta)$ denotes the galactocentric distance of a point along the line-of-sight integration path, with $r_\odot$ being the distance from the Sun to the Galactic Center, $s$ being the line-of-sight distance, and $\theta$ being the angle between the observation direction and the direction toward the Galactic Center. This expression encodes the standard line‑of‑sight integral of the squared DM density over the solid angle $\Delta\Omega$.

			To evaluate the Galactic‑Center production of boosted DM $\chi_2$ particles, we assume that $\chi_1$ follows the generalized Navarro–Frenk–White (gNFW) profile \cite{Arguelles:2019ouk}, which provides a flexible parametrization of the Milky‑Way halo, accommodating both cuspy and mildly cored central profiles. 
			
\begin{figure}
				\centering
				\includegraphics[width=0.5\textwidth]{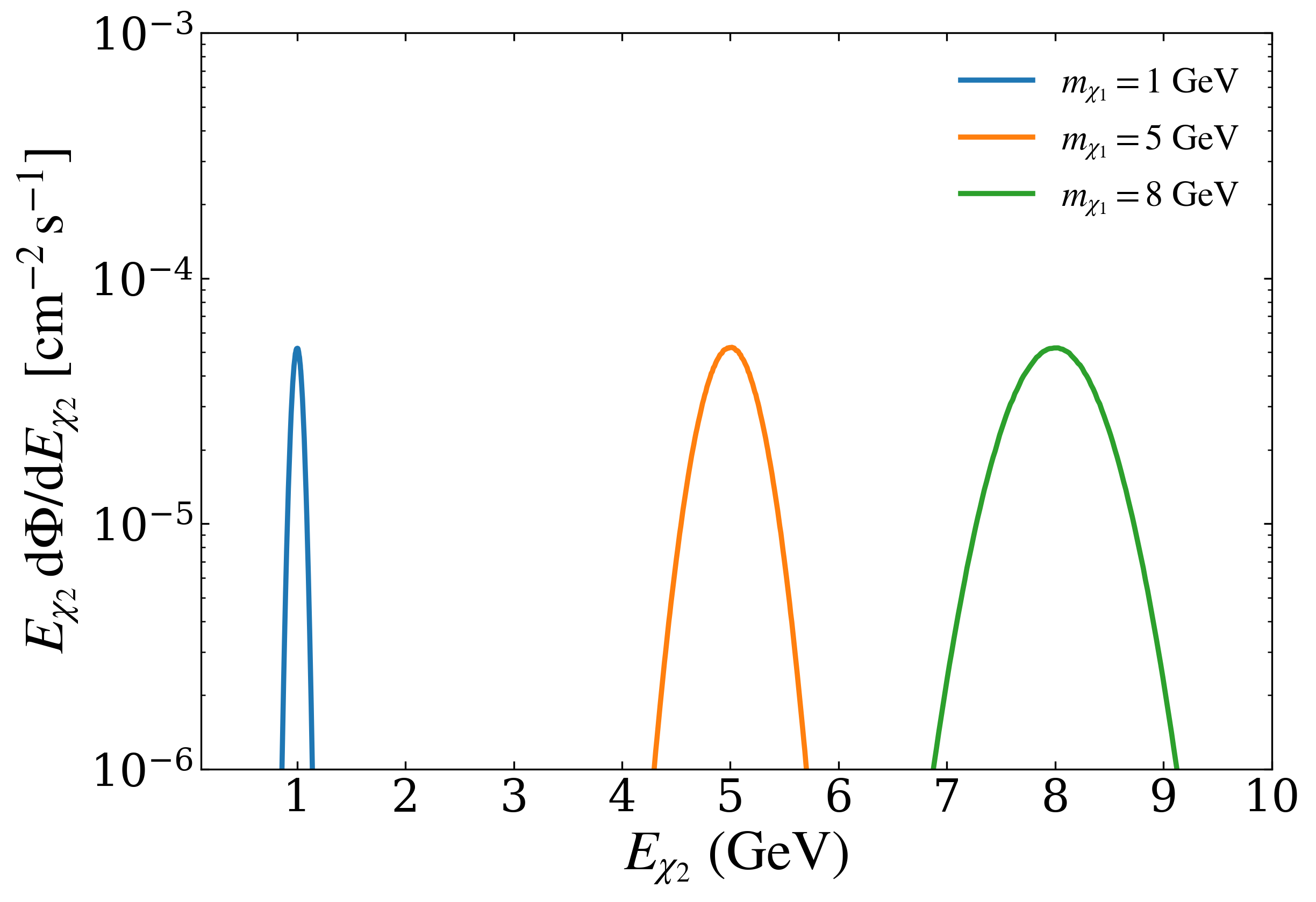}
				\vspace{-.2cm}
				\caption{
					Differential bare BDM flux of $\chi_2$ for different $m_{\chi_1}$ values: 
					the blue, orange, and green lines correspond to $m_{\chi_1}=1\,\mathrm{GeV}$, 
					$5\,\mathrm{GeV}$, and $8\,\mathrm{GeV}$, respectively.
				}
				\label{fig:Bare_Flux}
			\end{figure}

			The overall bare BDM flux is shown in Fig.\ref{fig:Bare_Flux}. Its peak, determined by Eq.\ref{eq:DES}, falls within the kinematic region where RES is significant for BDM produced from a GeV-scale $\chi_1$, making RES interactions non-negligible in the attenuation calculations.

			\section{DM-NUCLEUS RESONANCE CROSS-SECTION}
			In this section, we summarize the DM--nucleus resonance-scattering (RES) cross section mediated by a dark photon. The relevant interaction is
			\begin{equation}
				\mathcal{L}_{\mathrm{int}} = g\, \bar{\chi}\gamma^\mu \chi\, V^\prime_\mu + \epsilon e V^\prime_\mu J^\mu_{\mathrm{EM}},
			\end{equation}
			where \(g\) denotes the \(\chi\)-\(V'\) coupling, \(\epsilon\) is the kinetic mixing parameter, and \(e\) is the electromagnetic coupling. Through kinetic mixing, \(V'\) couples to the hadronic electromagnetic current \(J^\mu_{\mathrm{EM}}\). Although the neutron has zero net electric charge, \(J^\mu_{\mathrm{EM}}\) acts on its charged quark constituents and can induce \(\gamma^\ast n\to\Delta^0\). In the isospin-symmetric approximation, the electromagnetic \(N\to\Delta(1232)\) transition for \(p\to\Delta^+\) and \(n\to\Delta^0\) is described by the same reduced isovector matrix element. Since we treat the \(\Delta(1232)\) response inclusively, without resolving exclusive \(\pi N\) decay channels, we use the same elementary response for bound protons and neutrons.
			
			For an incoming DM particle with four-momentum \(k^\mu=(E_k,\vec{k})\) scattering into \(k^{\prime\mu}=(E_{k'},\vec{k}')\), the RES cross section can be written as
			\begin{equation}
				\frac{\mathrm{d} \sigma}{\mathrm{d}E_{k'}\mathrm{d} \Omega}
				=
				\frac{1}{16\pi ^2}
				\frac{\left\vert \vec{k}' \right\vert}{\left\vert \vec{k} \right\vert}
				\frac{g ^2 \epsilon ^2 e ^2}
				{(Q^2 + m _{V'} ^2)^2}
				L_{\chi}^{\mu\nu} W^A_{\mu \nu},
				\label{eq:res_xsec_basic}
			\end{equation}
			where \(q=k-k'\), \(Q^2\equiv -q^2>0\), and \(m_{V'}\) is the dark-photon mass. The factor \((Q^2+m_{V'}^2)^{-2}\) comes from the squared dark-photon propagator. The DM tensor is fixed by the vector current,
			\begin{equation}
				L_\chi^{\mu\nu}
				=
				\frac{1}{2}
				{\rm Tr}\left[
				(\slashed{k}'+m_\chi)\gamma^\mu
				(\slashed{k}+m_\chi)\gamma^\nu
				\right],
				\label{eq:dm_tensor}
			\end{equation}
			and is independent of the nuclear modeling. The nontrivial nuclear dynamics is contained in \(W_A^{\mu\nu}\), which describes resonance production inside the nucleus.

			\begin{figure}
				\centering
				\includegraphics[width=8cm,height=3.7cm]{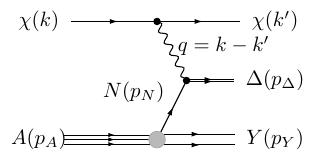}
				\caption{Diagrammatic sketch of resonant dark matter $\chi$  scattering off a bound nucleon $N$, $\chi(k) + A(p_A) \to \chi(k') + \Delta(p_\Delta)+Y(p_Y)$, exchanging a dark photon $V^\prime$.}
				\label{Fig:nuclear~FMD}
			\end{figure}
			
			In the impulse approximation(IA), the incoming DM particle scatters incoherently from one bound nucleon, while the remaining nucleons act as spectators. Accordingly, the nuclear current is written as a sum of single-nucleon currents,
			\begin{equation}
				J^\mu = \sum_{i=1}^{A} j_i^\mu .
				\label{eq:IA}
			\end{equation}
			
			\begin{figure*}
				\centering
				\includegraphics[width=0.95\textwidth]{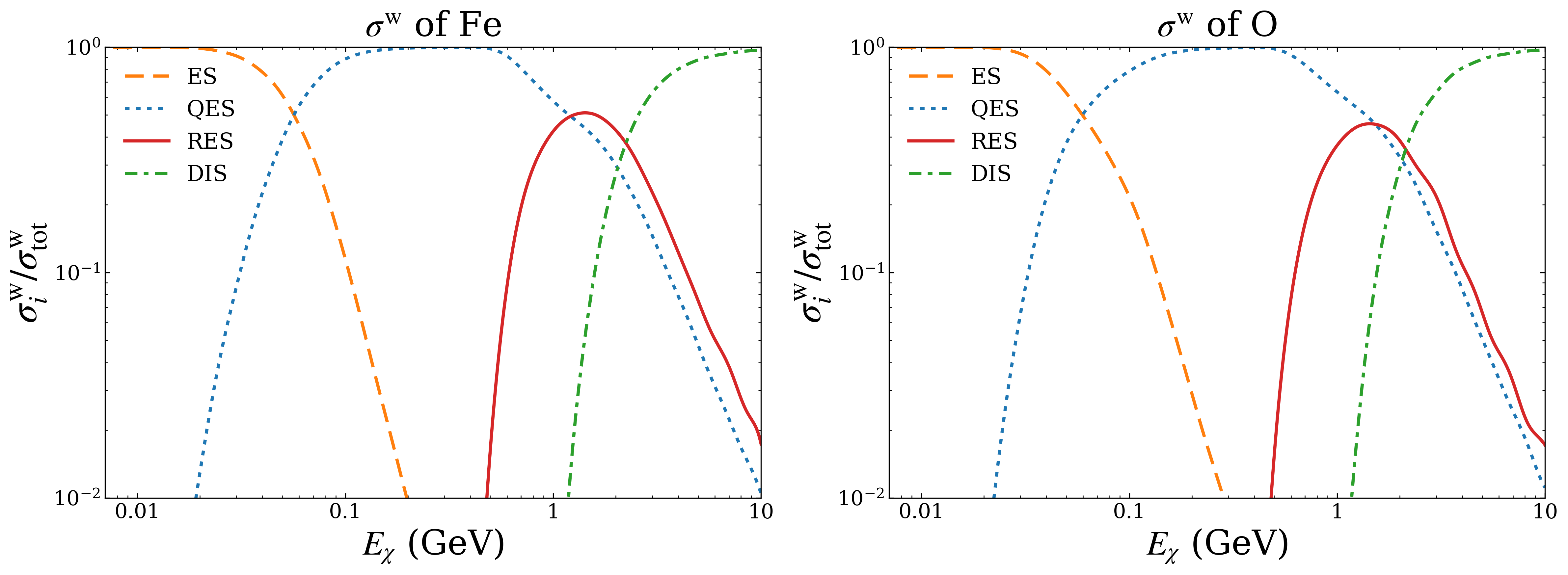}
				\vspace{-.2cm}
				\caption{
					Comparison of the effective energy-loss cross sections for Fe and O, with
					$m_V = 10~\mathrm{GeV}$ and $m_\chi = 1~\mathrm{MeV}$.
					The left and right columns correspond to Fe and O, respectively. The orange dashed, blue dotted, red solid, and green dash-dotted curves denote the elastic, quasi-elastic, resonant, and deep-inelastic contributions, respectively.
				}
				\label{fig:XS_comparison}
			\end{figure*}

			The nuclear tensor is then obtained by folding the elementary single-nucleon tensor with the nuclear spectral function,
			\begin{eqnarray}
				W^A_{\mu\nu}
				&=&
				\sum_i
				\int \mathrm{d}^3 p\, \mathrm{d}E\,
				\frac{m_N M_\Delta}{E_N E_\Delta}
				P(\vec{p},E)
				\nonumber\\
				&&\times
				W^i_{\mu\nu}
				\delta(m_N+E_k-E-E_{k'}-E_\Delta).
				\label{eq:res_nuclear_tensor}
			\end{eqnarray}
			Here \(m_N\) (\(E_N\)) and \(M_\Delta\) (\(E_\Delta\)) denote the mass (energy) of the initial nucleon \(N\) and the final baryon resonance \(\Delta\), respectively. The spectral function $P(\vec{p},E)$ gives the probability of finding a bound nucleon with momentum $\vec{p}$ and removal energy $E$, and is normalized according to 
			\begin{equation} 
				\int \mathrm{d}^3 p\,\mathrm{d}E\; P(\vec{p},E) =1.
			\end{equation}
			The elementary hadronic tensor \(W^i_{\mu\nu}\) is constructed from the standard electromagnetic \(N\to\Delta\) transition vertex \(\Gamma_{N\Delta}^{\alpha\mu}\), as used in pion-production and electroproduction studies~\cite{Gerasimov:2016zfr}.  In this work, we adopt the spectral function~\cite{Benhar:2005dj} as provided in NuWro~\cite{NuWro}. The energy-conserving delta function fixes the resonance kinematics after accounting for the off-shellness induced by nuclear binding.

			For an isoscalar target, and with a spectral function normalized per nucleon, the sum over struck nucleons gives an overall factor $A$ (mass number). The finite lifetime of the resonance state is incorporated by replacing the zero-width on-shell condition with a  Breit--Wigner distribution in the invariant mass \(W\equiv \sqrt{p_\Delta^2}\). In this work, we focus on the contribution of the first excited state of the nucleon, $\Delta(1232)$ with spin-3/2. Since its decay width, $\Gamma_{\Delta} \simeq 117$ MeV~\cite{ParticleDataGroup:2024cfk}, is much smaller than its mass ($\Gamma_\Delta \ll M_\Delta$), we keep the full $W$-dependence only in the Breit--Wigner factor and set $W\simeq M_\Delta$ in the slowly varying kinematic factors. After integrating over the azimuthal angle of the outgoing DM particle and using the energy-conserving delta function  to eliminate the polar angle of the initial nucleon momentum, the differential cross section can be rewritten as 
			\begin{eqnarray}
				&&\frac{\mathrm{d}\sigma}{\mathrm{d}E_{k^\prime} \mathrm{d}\cos{\theta}}=\frac{A}{8\pi }\frac{|\vec{k}'|}{|\vec{k}|}\frac{g ^2 \epsilon ^2 e ^2}{(Q ^2 +m _{V^\prime} ^2)^2} \nonumber \\
				&& ~~~~~~\times \int \mathrm{d}|\vec{p}|\, \mathrm{d}E\, \mathrm{d}\phi_p\, \mathrm{d}W\,
				\frac{|\vec{p}|}{|\vec{q}|}
				\frac{m_N M_\Delta}{E_N}
				P(|\vec{p}|,E)
				L_{\chi}^{\mu\nu}W_{\mu\nu}^i \nonumber\\
				&& ~~~~~~\times
				\frac{2M_\Delta \Gamma_\Delta}{\pi} 
				\frac{W}{(W^2 - M_\Delta^2)^2 + M_\Delta^2 \Gamma_\Delta^2},
				\label{eq:final}
			\end{eqnarray}
			Further details, including the explicit elementary \(N\to\Delta\) tensor, the Rarita--Schwinger spin-\(3/2\) projector, transition form factors, and the derivation of Eq.~\eqref{eq:final}, are provided in Appendix~\ref{app:resonance}.
			
			To compare the typical energy transfer associated with different scattering channels, we define the energy-transfer-weighted cross section
			\begin{equation}
				\sigma_{i,c}^{\rm w}(E_\chi)
				=
				\frac{1}{E_\chi}
				\int_0^{\omega_{\chi,i,c}^{\max}} d\omega_\chi\,
				\omega_\chi\,
				\frac{d\sigma_{i,c}}{d\omega_\chi}(E_\chi,\omega_\chi) ,
				\label{eq:sigma_weighted_def}
			\end{equation}
			where \(i\) labels the target nucleus and \(c=\mathrm{ES,QES,RES,DIS}\) labels the scattering channel, with \(\omega_\chi=E_\chi-E_\chi'\). For the ES, QES, and DIS channels, the differential cross sections are taken from Ref.~\cite{Su:2023zgr}. This quantity should be interpreted as an energy-transfer-weighted cross
			section rather than a total cross section. Figure~\ref{fig:XS_comparison} compares \(\sigma_{i,c}^{\rm w}\) for Fe and O for the heavy-mediator benchmark \(m_{V'}=10~\mathrm{GeV}\). In this case, the RES contribution is sizable for \(E_\chi\simeq1\)--\(2~\mathrm{GeV}\) in both Fe and O, indicating a non-negligible contribution to the corresponding energy-loss calculation.
			
			For a light mediator, the propagator factor \((Q^2+m_{V'}^2)^{-2}\) strongly enhances small-\(Q^2\) scattering. Compared with the elastic channel, this reduces the relative importance of larger-\(Q^2\) processes such as QES, RES, and DIS in DM--nucleus scattering. We therefore focus on the heavy-mediator benchmark, for which these inelastic channels can be more clearly assessed. In the following calculations, we take
			\(m_{V'}=10~\mathrm{GeV}\).
			
			In the attenuation analysis below, we include the RES contribution derived in this work together with the ES, QES, and DIS channels. For comparison between different mediator masses, we normalize the results to the reference momentum-independent DM--nucleon cross section
			\begin{equation}
				\bar{\sigma}_n
				=
				\frac{\epsilon^2 e^2 g^2 \mu_n^2}{\pi m_{V'}^4},
			\end{equation}
			where \(\mu_n\) is the DM--nucleon reduced mass.

			\section{Attenuation Effects}
			Direct DM detection experiments are typically located underground to suppress backgrounds from cosmic rays. As a result, the BDM flux reaching the detector differs from the flux at the Earth's surface in Eq.~(\ref{eq:source_flux}), owing to DM--nucleus scattering during propagation through the Earth. In this work, we estimate this attenuation effect using two simplified limiting descriptions: the straight-line model and the single-scattering model. The straight-line approximation is better motivated in the highly boosted regime, where the outgoing DM tends to be more forward-directed in the laboratory frame, particularly for scattering with individual partons. Its validity becomes less robust as $m_{\chi_2}$ approaches $m_{\chi_1}$ and the BDM becomes less relativistic, where sizable angular deflections may no longer be negligible.
				We therefore employ the straight-line and single-scattering models as two
				complementary limiting descriptions of BDM propagation through the Earth.
			
			Since the straight-line model is controlled by the energy-transfer-weighted cross sections discussed above, while the single-scattering model is governed by the ordinary total scattering cross sections, we present the straight-line treatment first and then the complementary single-scattering limit.

			\subsection{Straight-line model}
			In the straight-line model, interactions between BDM particles and nuclei are assumed not to deflect the BDM particles but only to reduce their kinetic energies. Consequently, the BDM particles remain on their original trajectories, while the energy spectrum is redistributed toward lower energies. The attenuated spectrum is then given by
			\begin{equation}
				\frac{\mathrm{d}\Phi_\chi^{\,z}}{\mathrm{d}E_\chi^{\,z}\, \mathrm{d}\Omega}
				=
				\frac{\mathrm{d}E_\chi}{\mathrm{d}E_\chi^{\,z}}\,
				\frac{\mathrm{d}\Phi_\chi}{\mathrm{d}E_\chi\, \mathrm{d}\Omega}
				\bigg|_{E_\chi = E_\chi(E_\chi^{\,z},\Omega)} ,
			\end{equation}
			where \(E_\chi(E_\chi^{\,z},\Omega)\) is the surface energy required to obtain \(E_\chi^{\,z}\) at the detector after propagation along direction \(\Omega\). For a direction-insensitive detector such as PandaX-4T, the observed energy spectrum is obtained by integrating over the incident solid angle,
			\begin{equation}
				\frac{\mathrm{d}\Phi_\chi^{\,z}}{\mathrm{d}E_\chi^{\,z}}
				=
				\int \mathrm{d}\Omega\,
				\frac{\mathrm{d}E_\chi}{\mathrm{d}E_\chi^{\,z}}\,
				\frac{\mathrm{d}\Phi_\chi}{\mathrm{d}E_\chi\, \mathrm{d}\Omega} \bigg|_{E_\chi = E_\chi(E_\chi^{\,z},\Omega)} .
			\end{equation}
			
			The relation between \(E_\chi\) and \(E_\chi^{\,z}\) is determined by the continuous energy loss accumulated along the trajectory, which is described by
			\begin{equation}
				\begin{aligned}
					\frac{dE_\chi(\ell)}{d\ell}
					&=
					- \sum_{i,c} n_i(r(\ell))
					\int_0^{\omega_{\chi,i,c}^{\max}} d\omega_\chi\,
					\omega_\chi\,
					\frac{d\sigma_{i,c}}{d\omega_\chi}
					\!\left(E_\chi(\ell),\omega_\chi\right) \\
					&=
					- E_\chi(\ell)
					\sum_{i,c} n_i(r(\ell))\,
					\sigma_{i,c}^{\rm w}\!\left(E_\chi(\ell)\right) .
					\label{eq:dEdz_general}
				\end{aligned}
			\end{equation}
			Here \(\ell\) denotes the path length along the trajectory, \(r(\ell)\) is the corresponding radial position inside the Earth, and \(n_i(r)\) is the number density of Earth component \(i\). The \(\sigma_{i,c}^{\rm w}\) is the energy-transfer-weighted cross section defined in Eq.~\ref{eq:sigma_weighted_def}. Solving Eq.~\ref{eq:dEdz_general} along each trajectory determines \(E_\chi(E_\chi^{\,z},\Omega)\) and the corresponding Jacobian entering the flux transformation.
			
			\subsection{Single–scattering model}
			
			In the single-scattering model, any interaction with an Earth nucleus is treated as removing the incoming DM particle from the incident flux. Therefore, only unscattered particles contribute to the attenuated flux, and their energy is unchanged, \(E_\chi^{\,z}=E_\chi\). The attenuated spectrum is therefore
			\begin{equation}
				\frac{\mathrm{d}\Phi_\chi^{\,z}}{\mathrm{d}E_\chi^{\,z}}
				=
				\int \mathrm{d}\Omega\,
				\mathcal{P}_{\mathrm{surv}}(E_\chi,\cos\theta)\,
				\frac{\mathrm{d}\Phi_\chi}{\mathrm{d}E_\chi\,\mathrm{d}\Omega}
				\bigg|_{E_\chi=E_\chi^{\,z}} .
				\label{eq:ss_flux}
			\end{equation}
			The factor \(\mathcal{P}_{\mathrm{surv}}\) is the probability that the DM particle reaches the detector without scattering, which is given by
			\begin{equation}
				\mathcal{P}_{\mathrm{surv}}(E_\chi,\cos\theta)
				=
				\exp\left[
				-\sum_i
				\frac{d_{\mathrm{eff},i}(\cos\theta)}
				{\bar{\lambda}_i(E_\chi)}
				\right],
				\label{eq:Psurv}
			\end{equation}
			
			Here \(\bar{\lambda}_i= \left[\sigma_i^{\mathrm{tot}}(E_\chi)\,\bar{n}_i\right]^{-1}\) is the mean free path for scattering off element \(i\), and \(\bar n_i=\int_0^{R_E}n_i(r)dr/R_E\) is the corresponding average number density. The effective propagation length \(d_{\mathrm{eff},i}\) encodes the density profile encountered along the incoming trajectory and is defined by
			\begin{equation}
				d_{\mathrm{eff},i}(\cos\theta)
				=
				\frac{1}{\bar n_i}
				\int_{\rm path} n_i(r)\,d\ell .
				\label{eq:deff_def}
			\end{equation}
			Using the spherical Earth density profile, this can be written approximately as
			\begin{equation}
				\begin{aligned}
					d_{\mathrm{eff},i}
					&~\approx
					\begin{cases}
						\displaystyle
						\int_{R_E\sin\theta}^{R_E}
						\frac{2r\,n_i(r)\,dr}
						{\bar{n}_i\sqrt{r^2-R_E^2\sin^2\theta}},
						& \displaystyle \theta \in \left[0,\frac{\pi}{2}\right],
						\\[2ex]
						\displaystyle
						\int_{R_E-h_D}^{R_E}
						\frac{n_i(r)}{\bar{n}_i}\,dr,
						& \displaystyle \theta \in \left[\frac{\pi}{2},\pi\right].
					\end{cases}
				\end{aligned}
			\end{equation}
			Here \(\theta\) is the angle between the direction from the Earth's center to the detector and the incoming DM direction, and \(h_D\) is the detector depth below the Earth's surface.

			\subsection{Attenuated flux at the detector}

			
			\begin{figure*}[ht!]
				\centering
				\includegraphics[width=1\textwidth]{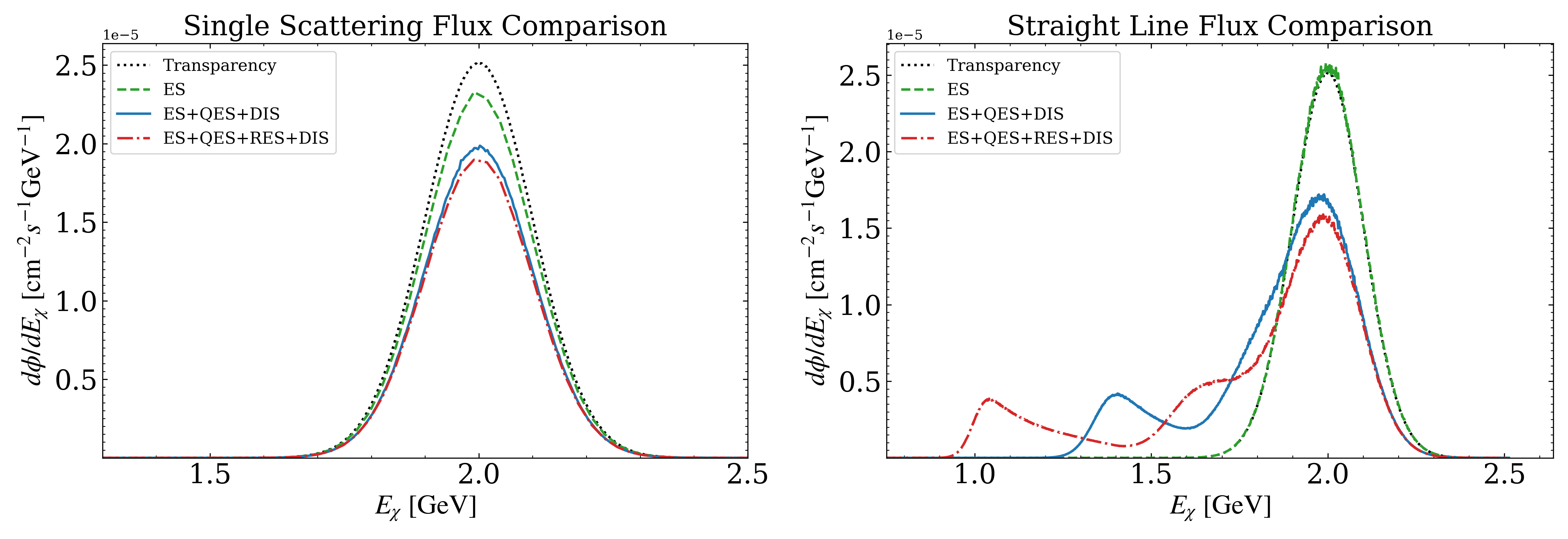}
				\vspace{-.2cm}
				\caption{Differential flux of boosted dark matter $\chi_2$ at the PandaX detector for $m_\chi = 1\,\mathrm{MeV}$, $m_V = 10\,\mathrm{GeV}$, $\sigma_0 = 0.05$, and $\bar{\sigma}_n=10^{-38}\mathrm{cm}^2$. 
					In each panel, four attenuation assumptions are compared: transparency (black), elastic only (green), elastic + quasi‑elastic + deep‑inelastic (blue), and the full calculation including the resonance contribution (red) .}
				\label{fig:flux_comparison}
			\end{figure*}
			The attenuation prescriptions described above enter the experimental analysis through the boosted-DM flux at the detector. Figure~\ref{fig:flux_comparison} shows this detector-level input for a representative parameter point, obtained by successively including different scattering channels in the propagation calculation while keeping the source flux fixed.
			
			The two attenuation limits modify the flux in qualitatively different ways. In the straight-line model, attenuation is controlled by the continuous energy-loss mapping between the surface and detector energies. The detector-level spectrum is then shaped by the shifted source flux together with the Jacobian of this mapping. Because the energy-loss rate is energy and channel dependent, different surface-energy intervals can be compressed into different detector-energy ranges, producing several local enhancements in the straight-line flux. In the single-scattering model, by contrast, propagation acts as an absorption factor: particles that interact are removed from the incident beam, and the spectrum is therefore suppressed without energy redistribution. The inclusion of the RES channel further enhances the redistribution effect in the straight-line calculation and the absorption effect in the single-scattering calculation around \(E_\chi\simeq1\)--\(2~\mathrm{GeV}\), where the RES contribution is sizable.
			
			We use these detector-level fluxes in the event-rate calculation and in deriving the experimental constraints below.
			
			\begin{figure*}
				\centering
				\includegraphics[width=0.95\textwidth]{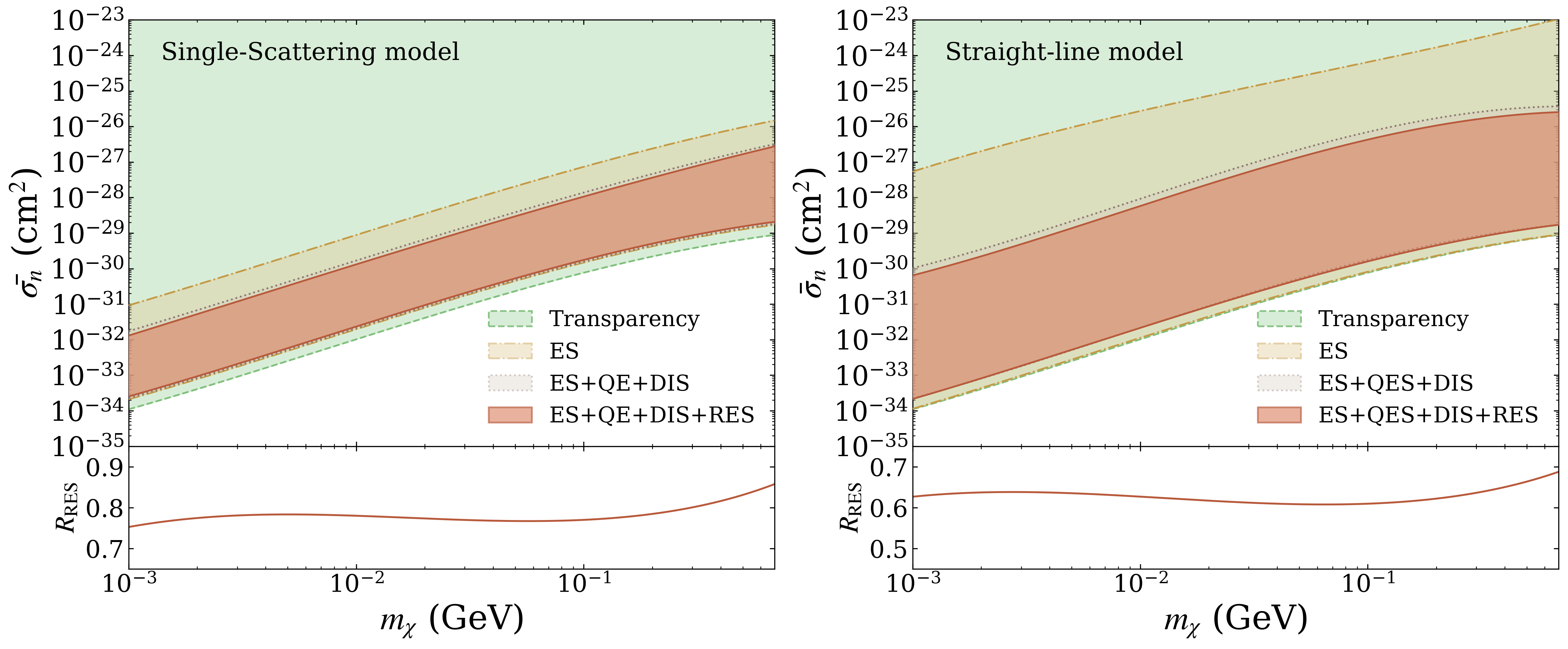}
				\vspace{-.2cm}
				\caption{
					Comparison of the 90\% C.L. exclusion limits derived from PandaX‑4T Run~1 data, obtained under the single‑scattering attenuation model (left) and the straight‑line attenuation model (right). 
					In both panels, the green region corresponds to the transparency limit (no attenuation), the yellow region includes elastic scattering only, the red region includes elastic + quasi‑elastic + deep‑inelastic processes, and the brown region further incorporates the resonant contribution. 
					Together, The lower subpanels show the ratio
					$R_{\rm RES}\equiv
					\bar{\sigma}^{\rm up}_n({\rm ES+QES+DIS+RES})/
					{\bar{\sigma}^{\rm up}_n({\rm ES+QES+DIS})}$
					where $\bar{\sigma}^{\rm up}_n$ denotes the upper boundary of the excluded region. 
				}
				\label{fig:exclusion_comparison}
			\end{figure*}
			\section{results}
			We use the PandaX-4T Run~1 data to derive the experimental constraints. The data correspond to an exposure of \(1~{\rm ton\cdot yr}\) of liquid xenon, with a nuclear-recoil analysis window of \(E_R=3\)--\(103~{\rm keV}\) \cite{PandaX:2024qfu}. Since this recoil window is at the keV scale, the detector response to the incident boosted-DM flux is dominated by elastic DM--xenon scattering. We therefore compute the signal event number using the elastic differential cross section for the detection process, \(\mathrm d\sigma_{ {\rm Xe}}^{\rm ES}/\mathrm dE_R\).

			For a detector-level BDM flux \(\mathrm{d}\Phi_\chi^z/\mathrm{d}E_\chi^z\), the expected number of nuclear-recoil events is
			\small
			\begin{equation}
				\mathcal{N}_{exp}= T_{\mathrm{Expo}} N_{T} \int_{E_{R}^{\min }}^{E_{R}^{\max }} \mathrm{d} E_{R} \int \epsilon\left(E_{R}\right) \frac{\mathrm{d} \Phi_{\chi}^{z}}{\mathrm{~d} E_{\chi}^{z}} \frac{\mathrm{~d} \sigma_{ {\rm Xe}}^{\rm ES}}{\mathrm{~d} E_{R}} \mathrm{~d} E_{\chi}^{z},
			\end{equation}
			\normalsize
			where \(T_{\rm Expo}N_T\) denotes the exposure factor and \(\epsilon(E_R)\) is the detection efficiency \cite{PandaX:2024med}. The recoil-energy integration range is \([E_R^{\min},E_R^{\max}]=[3,103]~{\rm keV}\). In the PandaX-4T Run~1 data set, 1373 events were observed in this window, consistent within \(1\sigma\) with the Standard Model expectation of \(1356\pm43\) events. We use these data to perform a likelihood-based analysis and derive the \(90\%\) confidence-level
			exclusion limits shown in Fig.~\ref{fig:exclusion_comparison}.
			
			The most important effect of Earth attenuation is the appearance of an upper boundary of the excluded region. In the transparency limit, increasing \(\bar\sigma_n\) monotonically enhances the event rate, so the exclusion extends to arbitrarily large cross sections within the range shown. Once attenuation is
			included, however, sufficiently large \(\bar\sigma_n\) suppresses the boosted-DM flux reaching the detector, and the predicted event rate drops below the experimental sensitivity. The excluded region is therefore bounded from above.
			
			The displacement of the upper boundary is mainly controlled by the attenuation of the flux near the peak of the incoming BDM spectrum, \(E_\chi\simeq2~\mathrm{GeV}\). Consequently, the relative shift induced by adding the QES, DIS, and RES channels approximately follows the increase in the effective attenuation strength around this energy. In the single-scattering treatment this strength is set by the ordinary total scattering cross section, whereas in the straight-line treatment it is set by the energy-transfer-weighted combination entering the continuous energy-loss equation. The sizable RES contribution around \(2~\mathrm{GeV}\) therefore leads to a visible lowering of the upper boundary once it is included.
			
			The lower subpanels provide a more direct visualization of the incremental effect of the RES channel on the upper boundary. We plot
			\begin{equation}
				R_{\rm RES}\equiv
				\frac{\bar{\sigma}^{\rm up}_n({\rm ES+QES+DIS+RES})}
				{\bar{\sigma}^{\rm up}_n({\rm ES+QES+DIS})},
			\end{equation}
			so that $R_{\rm RES}<1$ indicates that including RES lowers the upper boundary
				of the excluded region, thereby reducing the excluded parameter space
				and weakening the constraint. The ratio is smaller in the straight-line treatment than in the single-scattering treatment, implying a larger relative impact of RES in the continuous energy-loss description. This behavior is consistent with the fact that the straight-line model is governed by an energy-transfer-weighted cross section, which enhances the importance of channels with sizable energy transfer. The RES contribution is therefore more visible in this treatment than in the  total-cross-section used in the single-scattering model.
			
			The lower boundaries are less sensitive to attenuation than the upper boundaries, but they are not identical to the transparency limit. In the single-scattering treatment, as well as in the straight-line treatment once inelastic channels are included, attenuation of trajectories crossing the lower hemisphere reduces the effective flux and shifts the lower boundary upward by roughly a factor of two in the cases shown. The exception is the ES-only straight-line case: because elastic scattering transfers only a small fraction of the DM energy, the corresponding energy-transfer-weighted cross section is very small, and the resulting lower boundary remains close to the transparency curve.
			
			\section{Conclusion}
			In this work, we studied the RES contribution to DM--nucleus scattering for a Dirac DM particle coupled through a vector mediator. The RES channel lies between the QES and DIS regimes in energy transfer and describes the excitation of baryon resonances in the nucleus. For incoming DM energies of order GeV, we found that this channel gives a non-negligible contribution  in the heavy-mediator regime to both the total scattering cross section and the energy-transfer-weighted cross section.
			
			We then implemented the RES contribution in the Earth-attenuation calculation and assessed its impact on underground xenon searches. Applying the analysis to the PandaX-4T Run~1 data, we found that the RES contribution at GeV energies modifies the attenuated boosted-DM flux at the detector and lowers the attenuation induced upper boundary of the excluded region. This application illustrates the broader phenomenological relevance of the RES channel. In GeV-scale DM--nucleus scattering, RES can provide a sizable contribution in the baryon-resonance region and should be included explicitly.
			
			\section*{acknowledgments}
			This work is supported by 
			the NNSFC under Grant No.~12275134, No.~12335005, and by the PI Research Fund from Henan Normal University under Grant No. 5101029470335. L.S. is supported by the Alexander von Humboldt Foundation. L.W. acknowledges
			support from the State Key Laboratory of Dark Matter Physics.

			\appendix
			\section{Derivation of the resonant nuclear response}
			\label{app:resonance}
			The RES contribution corresponds to the excitation of a bound nucleon into a baryon-resonance state, dominantly the \(\Delta(1232)\), induced by the momentum transfer from the dark-sector current. The corresponding differential cross section can be written as
			\begin{eqnarray}
				d \sigma&=& \frac{1}{64\pi ^3m_A}\frac{\left\vert \vec{k^\prime} \right\vert}{\left\vert \vec{k} \right\vert}dE _{k ^{\prime}}d\Omega \left( \prod _f \frac{d ^3   p _f ^{\prime}}{ (2\pi)^3 2 E _f}\right) \nonumber \\
				&& \times (2\pi)^4 \delta ^4  (p_X+k ^{\prime}-p_A -k)\left\vert M \right\vert^2
			\end{eqnarray}
			where $p_X$ denotes the total final-state hadronic four-momentum. We define the four-momentum transfer as \[q^\mu=k^\mu-k^{\prime\mu},\qquad Q^2=-q^2>0,\] where \(k^\mu\) and \(k^{\prime\mu}\) are the incoming and outgoing DM four-momenta. After separating the dark-sector tensor from the nuclear response, the double differential cross section takes the compact form
			\begin{equation} 
				\frac{\mathrm{d} \sigma}{\mathrm{d}E_{k ^{\prime}}\mathrm{d} \Omega}=\frac{1}{16\pi ^2}\frac{\left\vert \vec{k ^{\prime}} \right\vert}{\left\vert \vec{k} \right\vert}\frac{g ^2 \epsilon ^2 e ^2}{(Q^2 + m _{V^\prime} ^2)^2} L _{\mu\nu} W_A ^ {\mu \nu},
			\end{equation}
			where the dark-sector tensor $L_{\mu\nu}$ can be expressed as:
			\begin{equation}
				L^{\mu\nu}=2g^{\mu\nu}(m_\chi^2-k.k^\prime)+2(k^\mu k^{\prime\nu}+k^{\prime\mu }k^\nu)
			\end{equation}
			Within the impulse approximation of Eq.~\eqref{eq:IA}, the nuclear hadronic tensor $W_A^{\mu\nu}$ can be written as a weighted integral of the single-nucleon tensor $W_i^{\mu\nu}$ with the nucleon spectral function $P(p, E)$ as
			\begin{equation}
				\begin{aligned}
					W_A^{\mu\nu}=&\sum _i\int d ^3 p dE  \frac{m_NM_\Delta}{E_N E_{\Delta}}P(p,E) \\
					&~~~~\times W ^{\mu\nu}_i(p,p_\Delta) \delta(m_N+E_k-E-E_{k ^{\prime}}-E_\Delta)
				\end{aligned}
			\end{equation}
			where \(i\) labels the struck bound nucleon in the IA sum, spectral function $P(p,E)$ as
			\small
			\begin{equation} 
				P(p,E)=\sum _Y\left\vert \bra{A}\ket{Y,-p_Y;N,p_Y} \right\vert^2 \delta(E-E_Y+E_A-m_N)
			\end{equation}
			\normalsize
			and $W_i^{\mu\nu}$ can be derived following the parametrization of Ref.~\cite{Gerasimov:2016zfr} as:
			\small
			\begin{equation}
				\begin{aligned}
					W_i^{\mu\nu}(p,p_\Delta)&=\frac{(M_\Delta+m_N)^2}{16m_N^3 M_\Delta}((M_\Delta-m_N)^2-\tilde{q}^2)\\
					&\times\left[(G_M^{\star2}(\tilde{Q}^2)+3G_E^{\star2}(\tilde{Q}^2))\left(-g^{\mu\nu}+\frac{\tilde{q}^\mu\tilde{q}^\nu}{\tilde{q}^2}+\frac{\tilde{P}^\mu\tilde{P}^\nu}{\tilde{P}^2}\right)\right.\\
					&~~~\left.-\frac{\tilde{q}^2}{M_\Delta^2}G_C^{\star2}(\tilde{Q}^2)\frac{\tilde{P}^\mu\tilde{P}^\nu}{\tilde{P}^2}\right]
				\end{aligned}
			\end{equation}
			\normalsize
			where $\tilde{q}=p_\Delta-p$, $\tilde{Q}^2=-\tilde{q}^2$, and
			\begin{equation}
				\tilde{P}^\mu=p^\mu+p_\Delta^\mu-\frac{(p+p_\Delta)\cdot\tilde{q}}{\tilde{q}^2}\tilde{q}^\mu
			\end{equation}
			The empirical form factors can be taken from MAID2007 analysis\cite{Drechsel:2007if,Tiator:2011pw} as
			\begin{equation}
				G_{M,E,C}^{*}(Q^2) = \left( \frac{\sqrt{Q^2+(M_\Delta+m_N)^2}}{M_N + M_\Delta} \right)
				G_{M,E,C}^{*\mathrm{Ash}}(Q^2), 
			\end{equation}
			where the Ash form factors $G_{M,E,C}^{*\mathrm{Ash}}$ are parametrized as
			\begin{align}
				G_{M}^{*\mathrm{Ash}}(Q^2) &= 3.00\,(1 + 0.01\,Q^2)\, e^{-0.23\,Q^2}\, G_D(Q^2), \nonumber \\
				G_{E}^{*\mathrm{Ash}}(Q^2) &= 0.064\,(1 - 0.021\,Q^2)\, e^{-0.16\,Q^2}\, G_D(Q^2), \nonumber \\
				G_{C}^{*\mathrm{Ash}}(Q^2) &= 0.124\,
				\frac{(1 + 0.120\,Q^2)}{1 + 4.9\,Q^2/4m_N^2}
				\left( \frac{4M_\Delta^2}{M_\Delta^2 - m_N^2} \right) \nonumber \\
				&\qquad \times e^{-0.23\,Q^2}\, G_D(Q^2), 
			\end{align}
			for $Q^2$ in $\mathrm{GeV}^2$, the dipole form factor is expressed as $G_D(Q^2) = 1/(1 + Q^2/0.71)^2$.
			
			Collecting these results, in the zero-width approximation the double differential cross section takes the form
			\begin{equation}
				\begin{aligned}
					\frac{d \sigma}{dE_{k'}d \Omega}
					= &\frac{1}{16\pi ^2}\frac{|\vec{k}'|}{|\vec{k}|}
					\frac{g ^2 \epsilon ^2 e ^2}{(Q ^2 +m _{V^\prime} ^2)^2} \\
					&\quad\times \sum_i\int d ^3 p\, dE\,
					\frac{m_NM_\Delta}{E_N E_{\Delta}}
					P(p,E) \\
					&\quad\times L _{\mu\nu} W ^{\mu\nu}_i\,
					\delta(m_N+E_k-E-E_{k'}-E_\Delta) 
				\end{aligned}
			\end{equation}
			Here \(E_\Delta=\sqrt{|\vec p+\vec q|^2+M_\Delta^2}\). We choose the
			\(z\)-axis along \(\vec q\), so that
			\begin{equation} 
				d^3p=|\vec p|^2 d|\vec p|\,d\cos\theta_p\,d\phi_p ,
			\end{equation}
			where \(\theta_p\) is the polar angle between \(\vec p\) and \(\vec q\), and \(\phi_p\) is the corresponding azimuthal angle.
			
			To include the finite width of the resonance, we insert an integral over the invariant mass \(W\) of the final hadronic state. In the zero-width limit the resonance contribution is proportional to \(\delta(W-M_\Delta)\), which is replaced by the normalized Breit--Wigner distribution
			\begin{equation}
				\begin{aligned}
					\delta(W-M_\Delta)
					&\longrightarrow
					\frac{2M_\Delta \Gamma_\Delta}{\pi}
					\frac{W}{(W^2-M_\Delta^2)^2+M_\Delta^2\Gamma_\Delta^2}.
					\label{eq:BW_replacement}
				\end{aligned}
			\end{equation}
			
			Following Ref.~\cite{Gerasimov:2016zfr}, we neglect the difference between \(W\) and \(M_\Delta\) in the slowly varying kinematic factors and in the single-nucleon tensor, while keeping the full \(W\)-dependence in the Breit--Wigner factor.
			
			Utilizing 
			\begin{equation} 
				E_\Delta=\sqrt{M_{\Delta}^2+\left\vert \vec{p} \right\vert^2+ \left\vert \vec{q} \right\vert^2 +2 \left\vert \vec{p} \right\vert \left\vert \vec{q} \right\vert \cos{\theta_p}}
			\end{equation}
			we rewrite the $\delta$ function as
			\begin{equation} 
				\delta(m_N+E_k-E-E_{k^\prime}-E_\Delta)=\frac{E_{\Delta}}{\left\vert \vec{p} \right\vert \left\vert \vec{q} \right\vert} \delta(\cos{\theta _p}-\cos{\theta _p ^0})
			\end{equation}
			Using the energy-conservation delta function to perform the \(\cos\theta_p\) integral, we obtain the double differential cross section as 
			\begin{eqnarray}
				&&\frac{d\sigma}{dE_{k^\prime} d\cos{\theta}}=\frac{A}{8\pi }\frac{\left\vert \vec{k ^{\prime}} \right\vert}{\left\vert \vec{k} \right\vert }\frac{g ^2 \epsilon ^2 e ^2}{(Q ^2 +m _{V^\prime} ^2)^2} \nonumber \\
				&& ~~~~~~\times \int d|\vec{p}|\, dE\, d\phi_p\, dW\,
				\frac{|\vec{p}|}{|\vec{q}|}
				\frac{m_N M_\Delta}{E_N}
				P(|\vec{p}|,E)
				L_{\mu\nu}W_i^{\mu\nu} \nonumber\\
				&& ~~~~~~\times
				\frac{2M_\Delta \Gamma_\Delta}{\pi} 
				\frac{W}{(W^2 - M_\Delta^2)^2 + M_\Delta^2 \Gamma_\Delta^2}.
			\end{eqnarray}

			\bibliography{refs}
			
		\end{document}